\documentclass[%
reprint,
nofootinbib,
 amsmath,amssymb,
 aps,
 prd,
floatfix,
]{revtex4-1}

\usepackage[T1]{fontenc}
\usepackage[utf8]{inputenc}
\usepackage{amsmath,amssymb,bm,mathtools}
\usepackage{xcolor}
\usepackage[normalem]{ulem}
\DeclareRobustCommand{\added}[1]{{\color{blue}#1}}
\DeclareRobustCommand{\deleted}[1]{{\color{red}\sout{#1}}}

\DeclareRobustCommand{\deletedKerr}[1]{\ifhmode\allowbreak\fi{%
  \renewcommand{\added}[1]{##1}%
  \renewcommand{\deleted}[1]{##1}%
  \let\kerrcite\cite
  \let\kerreqref\eqref
  \let\kerrref\ref
  \renewcommand{\cite}[1]{\mbox{\kerrcite{##1}}}%
  \renewcommand{\eqref}[1]{\mbox{\kerreqref{##1}}}%
  \renewcommand{\ref}[1]{\mbox{\kerrref{##1}}}%
  \hypersetup{citecolor=red,linkcolor=red}%
  \color{red}\sout{#1}}\ifhmode\allowbreak\fi}

\usepackage{pgfplots}
\usepackage{microtype}
\usepackage{graphicx,tikz}
\usepackage{dcolumn}
\usepackage{hyperref}
\usepackage{natbib}
\hypersetup{
    colorlinks=true,
    linktoc=page,    
    linkcolor=blue,
    citecolor=blue,
    urlcolor=blue
}

\newcommand{\TE}{T_{\mathrm E}}
\renewcommand{\TH}{T_{\mathrm H}}
\newcommand{\GN}{G_{\mathrm N}}
\newcommand{\IE}{I_{\mathrm E}}
\newcommand{\Fgen}{F_{\mathrm{gen}}}
\newcommand{\Fcl}{F_{\mathrm{cl}}}
\newcommand{\Fav}{\bar F}
\newcommand{\Ftherm}{F_{\mathrm{therm}}}
\newcommand{\Geff}{\Gamma_{\mathrm{eff}}}
\newcommand{\dd}{\mathop{}\!\mathrm{d}}
\newtheorem{proposition}{Proposition}
\newenvironment{propositionproof}
 {\par\noindent\textit{Proof.}\ }
 {\par}

\begin{document}
\raggedbottom

\title{Saddle-Order Universality in Ensemble-Averaged Black Hole Thermodynamics}

\author{Ankit Anand}
\email{anand@iitk.ac.in}
\affiliation{Department of Physics, Indian Institute of Technology Kanpur, Kanpur 208016, India.}

\author{Peng Cheng}
\email{p.cheng.nl@outlook.com}
\affiliation{Center for Joint Quantum Studies and Department of Physics, School of Science, Tianjin University, Tianjin 300350, China.}

\begin{abstract}
We establish a saddle-order proposition for the ensemble-averaged generalized free energy in reduced Euclidean black hole ensembles. Under regularity and stability assumptions, the leading fluctuation contribution is determined by the order of the first stabilizing term around the dominant saddle. An ordinary critical point is governed by a quartic soft mode, giving a quarter-temperature contribution instead of the familiar half-temperature Gaussian result. We verify this analytically for four-dimensional RN-AdS and Kerr-AdS black holes and map their thermodynamic controls to a common quartic normal form, whose scaling function connects the critical and Gaussian regimes along the zero-field one-well direction. The proposition also extends to coupled collective modes with weighted-homogeneous leading potentials. These results provide a unified description of reduced-ensemble fluctuations through black hole criticality.
\end{abstract}

\maketitle

\section{Introduction}

Black hole thermodynamics links classical gravitation to quantum theory by identifying horizon area with entropy and surface gravity with temperature \cite{Bekenstein1973,Hawking1975,Hawking1976}.
Bekenstein's entropy proposal and Hawking's radiation calculation turn the laws of black hole mechanics into thermodynamic relations and make stationary horizons genuine thermal systems.
They also provide a natural setting in which the gravitational path integral acquires a direct thermodynamic interpretation.
In the Euclidean formulation, periodic imaginary time supplies the thermal boundary condition, smoothness at the horizon fixes the Hawking period, and the on-shell action gives the leading equilibrium potential \cite{HartleHawking1976,GibbonsHawking1977,HawkingPage1983}.
Canonical constructions further show that the boundary data and ensemble are essential for stability \cite{York1986,WhitingYork1988,BradenBrownWhitingYork1990}.

Off-shell Euclidean treatments permit a conical defect when the thermal period is varied independently of the horizon geometry, with the opening angle conjugate to the horizon area~\cite{CarlipTeitelboim1995,FrolovIsraelSolodukhin1996}.
Related free-energy landscape descriptions study Hawking-Page and RN-AdS transitions through the generalized free energy of conically singular geometries~\cite{LiWang2020,LiZhangWang2020,LiWang2022}.
Building on this off-shell viewpoint, a reduced ensemble was introduced that integrates over a continuous family of geometries at fixed boundary temperature and conserved charges \cite{ChengLiuWei2025}.
A collective coordinate, such as the horizon radius, labels this family, and the smooth black hole is its stationary member.
This restriction of the gravitational path integral assigns a probability distribution to the off-shell free-energy landscape.

For Schwarzschild-AdS, RN-AdS, and Kerr-AdS, the normalized average of the generalized free energy was found to be
\begin{equation}
 \Fav=\Fcl+\frac{\TE}{2}+\cdots .
 \label{eq:oldhalf}
\end{equation}
Here $\TE$ is the ensemble temperature and $\Fcl$ is the generalized free energy at the classical saddle.
The Kerr-AdS analysis traced the coefficient to a Gaussian integral and argued that it is independent of the black hole equation of state \cite{ChengPanXuYang2025}.
We therefore ask which part of the fluctuation correction is fixed by the order of the dominant Euclidean saddle, and how the answer changes when the Gaussian saddle becomes degenerate.

The Gaussian expansion assumes that the Hessian at the dominant saddle is nondegenerate.
At thermodynamic criticality, one of its eigenvalues vanishes, defining a soft collective mode.
The Gaussian one-loop approximation to the reduced effective action is singular in this direction \cite{ChengLiuYu2026}.
Euclidean fluctuation modes have long provided a related diagnostic of black hole thermodynamic stability \cite{GrossPerryYaffe1982,Prestidge2000,MonteiroSantos2009}.
The local potential of the critical direction is instead quartic at an ordinary critical point and has the $A_3$ cusp structure found in off-shell analyses \cite{HaoWang2026}.
The same quartic integral appears in homogeneous finite-size scaling \cite{BrezinZinnJustin1985}.

The vanishing curvature changes the scale of the fluctuations.
Away from criticality, the width of the collective coordinate is proportional to $g^{1/2}$, where $g\ll1$ is the dimensionless semiclassical parameter for which the reduced Euclidean action scales as $\IE=\mathcal O(g^{-1})$; for RN-AdS, $g=\GN/L^2$ with $L$ the AdS radius.
At criticality the quadratic restoring force vanishes, and the quartic term gives a width proportional to $g^{1/4}$.
The semiclassical expansion is therefore nonuniform as $g\to0$ and the critical point is approached, requiring a uniform quartic integral to connect the Gaussian and critical regimes.

Equation~\eqref{eq:oldhalf} concerns the normalized first moment of the off-shell generalized free energy.
This observable is distinct from the thermodynamic potential $-\TE\ln Z$ obtained from the unnormalized reduced partition function $Z$.
For a unique stable interior saddle with a regular nonzero measure density, the saddle order fixes the leading normalized fluctuation contribution.
A coordinate-independent normalization cancels from a probability average, but it contributes to the logarithm of an unnormalized partition function.
In particular, any power of $g$ in that normalization contributes to the logarithmic coefficient along with the shrinking local integration volume.
Universality of the normalized moment therefore does not by itself fix the logarithmic correction to the full gravitational partition function.

We formulate the answer as a saddle-order proposition for the reduced ensemble.
Under these regularity assumptions, a leading potential of even order $p$ gives a single-mode contribution $\TE/p$ to $\Fav-\Fcl$.
The Gaussian case $p=2$ yields $\TE/2$, while an ordinary critical mode with $p=4$ contributes $T_c/4$ at the critical temperature $T_c$.
The integral identity is a form of generalized equipartition.
The gravitational calculation must establish the reduced measure and saddle order with consistent semiclassical scaling, and determine how physical control parameters unfold the saddle.

We establish a saddle-order universality for fluctuation corrections in reduced Euclidean black hole ensembles and demonstrate it explicitly in RN-AdS and Kerr-AdS thermodynamics.
Section~\ref{sec:ensemble} defines the reduced ensemble and proves the saddle-order proposition, while Subsec.~\ref{sec:multidimensional} extends it to coupled collective directions with a weighted homogeneous leading potential.
Sections~\ref{sec:rn} and \ref{sec:kerr} provide analytic realizations in RN-AdS and Kerr-AdS, where ordinary thermodynamic criticality produces a quartic saddle and the corresponding quarter-temperature contribution. They also map the physical control parameters to the same quartic normal form and describe the crossover to the Gaussian regime.
Section~\ref{sec:discussion} discusses the physical scope of the result, while Appendix~\ref{app:general-polynomial} examines how polynomial couplings determine the dominant semiclassical scaling.

\section{Reduced ensemble and saddle-order universality}
\label{sec:ensemble}

\subsection{Normalized observable and measure}

Let $q$ label a smooth off-shell black hole family at fixed boundary data and ensemble temperature $\TE=\beta^{-1}$.
Its reduced action is $\IE(q)=\beta\Fgen(q)$.
Write its measure as
\begin{equation}
 \dd\nu_g(q)=\mathcal N(g)\rho(q,g)\dd q,
 \label{eq:measurefactor}
\end{equation}
where the normalization $\mathcal N(g)$ is independent of $q$.
A source that scales the reduced action defines
\begin{equation}
 Z(s)=\int_{\mathcal D}\dd\nu_g(q)e^{-s\IE(q)}.
 \label{eq:sourceZ}
\end{equation}
The domain $\mathcal D$ is the physical range of the collective coordinate.
For an integral over the horizon radius, it begins at the extremal configuration when one exists and extends through the nonextremal branch.
The density $\rho$ contains the Jacobian associated with the chosen reduced measure.
Its smoothness near the dominant saddle is part of the local approximation, but its behavior at an endpoint can still determine whether the saddle is an interior one.

At $s=1$, the normalized weight and the generalized free energy average are
\begin{align}
 \dd P(q)&=\frac{\dd\nu_g(q)e^{-\IE(q)}}{Z(1)}, \notag\\
 \Fav&=\int_{\mathcal D}\dd P(q)\Fgen(q)
 =-\TE\left.\frac{\dd}{\dd s}\ln Z(s)\right|_{s=1}.
 \label{eq:sourceFav}
\end{align}
Thus $\Fav$ is a coordinate-independent source response within the reduced ensemble.
The source derivative follows directly from $\IE=\beta\Fgen$ and gives 
\begin{equation}
	-\frac{\dd\ln Z}{\dd s}=\langle\IE\rangle_s\,
\end{equation}
and we have  $\Fav=\TE\langle\IE\rangle$ at $s=1$.
At $s=1$, the classical part is the action at the dominant smooth geometry and the remaining term measures the local off-shell fluctuations retained by the reduced integral.
The equilibrium thermodynamic potential constructed from the same integral is
\begin{equation}
 \Ftherm=-\TE\ln Z(1).
 \label{eq:Ftherm}
\end{equation}
An overall factor in the measure cancels from Eq.~\eqref{eq:sourceFav} but remains in Eq.~\eqref{eq:Ftherm}.
If $\mathcal N(g)=\mathcal N_0g^\alpha[1+\mathcal O(g^\epsilon)]$ for some $\epsilon>0$, the complete logarithmic coefficient in $-\ln Z$ contains $\alpha$ in addition to the power supplied by the local saddle integral.

The normalized moment is also unchanged by an ordinary coordinate transformation.
If $q=q(x)$ has a finite nonzero Jacobian at the saddle, then
\begin{equation}
\rho_x(x,g)=\rho(q(x),g)\left|\frac{\dd q}{\dd x}\right|	
\end{equation}
A transformation that depends on $g$ can instead move powers of the semiclassical parameter between the local density and the action.
The asymptotic coordinate below is therefore dimensionless and held fixed as $g\to0$.

\subsection{Saddle-order proposition}

Consider a semiclassical family for which the reduced Euclidean action scales as $\mathcal O(g^{-1})$ and the dominant saddle remains in the interior of $\mathcal D$.
The local collective coordinate is taken to be independent of $g$, so that the semiclassical scaling is carried by the shrinking saddle region rather than by a $g$-dependent redefinition of the coordinate.
For a stable saddle on a real contour, the first nonvanishing term in the local expansion must have even order with a positive coefficient.
Under these assumptions, the same local rescaling controls both the leading normalized moment and the prefactor of the corresponding partition function.
This gives the following saddle-order proposition.

\begin{proposition}
\label{prop:saddleorder}

Let $q_0$ be a unique stable interior minimum on a real integration contour, and choose $\delta=q-q_0$ independently of the semiclassical parameter $g$.
Suppose that near $q_0$
\begin{align}
 \IE(q_0+\delta)
 &=I_0+\frac{1}{g}
 \left[
 \widehat a_p\delta^p
 +\widehat a_{p+1}\delta^{p+1}
 +\mathcal O(\delta^{p+2})
 \right],
 \notag\\
 \rho(q_0+\delta,g)
 &=\rho_0+\rho_1\delta+\mathcal O(\delta^2),
 \label{eq:orderp}
\end{align}
where $p$ is even, $\widehat a_p>0$, and $\rho_0>0$.

Let $Z_{\mathrm{loc}}$ denote the contribution of this saddle to the reduced partition function.
Then
\begin{align}
 \Fav-\Fcl
 &=\frac{\TE}{p}
 +\mathcal O(\TE g^{2/p}),
 \qquad
 \Fcl=\frac{I_0}{\beta},
 \label{eq:mainlawerror}\\
 Z_{\mathrm{loc}}
 &=e^{-I_0}\mathcal N(g)\rho_0 C_p
 \left(\frac{g}{\widehat a_p}\right)^{1/p}
 \left[1+\mathcal O(g^{2/p})\right],
 \label{eq:prefactor}
\end{align}
where
\begin{equation}
 C_p
 \equiv
 \int_{-\infty}^{\infty}\dd x\,e^{-x^p}
 =
 \frac{2}{p}\Gamma\!\left(\frac{1}{p}\right).
 \label{eq:cpdef}
\end{equation}

\end{proposition}

\begin{propositionproof}
Set $\delta=(g/\widehat a_p)^{1/p}x$.
The leading weight becomes $e^{-x^p}$, and the integration limits may be extended to the real line because the minimum remains a finite distance from the boundary.
The terms of order $g^{1/p}$ arise from the coefficient of $\delta^{p+1}$ and the first derivative of the density.
They are odd in $x$ and therefore integrate to zero.
The terms from $\delta^{p+2}$, the second derivative of the density, and products of the odd corrections first contribute at order $g^{2/p}$.
Direct evaluation of the leading integral then gives Eq.~\eqref{eq:prefactor}.

The normalized moment follows from the same scaled integral.
Integration by parts gives
\begin{equation}
 0=\int_{-\infty}^{\infty}\dd x\,
 \frac{\dd}{\dd x}\left(xe^{-x^p}\right),
 \qquad
 \langle x^p\rangle_0=\frac{1}{p}.
 \label{eq:virial}
\end{equation}
The boundary term vanishes because $p$ is even and positive.
In the original coordinate, Eq.~\eqref{eq:virial} states that the mean leading potential $\widehat a_p\delta^p/g$ is $1/p$.
The order-$g^{1/p}$ corrections are again odd and vanish in the normalized moment, so the first local correction is of order $g^{2/p}$.

Equivalently, the source-dependent local integral contains the factor $e^{-sI_0}s^{-1/p}$.
Applying Eq.~\eqref{eq:sourceFav} gives
\begin{equation}
 \TE\left(I_0+\frac{1}{p}\right)
 +\mathcal O(\TE g^{2/p}).
\end{equation}
Subtracting $\Fcl=\TE I_0$ proves Eq.~\eqref{eq:mainlawerror}.
The common factor $\mathcal N(g)\rho_0$ cancels from the normalized moment.

\end{propositionproof}

If the reduced action or density contains an additional explicit semiclassical correction of order $g^\kappa$, the remainder $g^{2/p}$ in Eqs.~\eqref{eq:mainlawerror} and \eqref{eq:prefactor} is replaced by $g^\eta$, where
\begin{equation}
 \eta=\min\left(\frac{2}{p},\kappa\right).
\end{equation}

The requirement of a unique interior minimum makes the result genuinely local.
Other saddles are exponentially suppressed when their actions remain separated from the dominant saddle by a positive amount of order $g^{-1}$.
The extension of the local integration limits to the real line then introduces only an exponentially small error.
Degenerate global minima instead have comparable weights and must be summed before a normalized moment is formed.

The proposition separates the normalized contribution from the full partition-function prefactor.
For $\mathcal N(g)=\mathcal N_0g^\alpha[1+\mathcal O(g^\epsilon)]$ with $\epsilon>0$, Eq.~\eqref{eq:prefactor} implies
\begin{align}
 -\ln Z_{\mathrm{loc}}-I_0
 ={}&
 \left(\alpha+\frac{1}{p}\right)\ln\!\left(\frac{1}{g}\right)
 +\frac{1}{p}\ln\widehat a_p \notag\\
 &-\ln(\mathcal N_0\rho_0C_p)
 +\mathcal O\!\left(g^{\min(2/p,\epsilon)}\right).
 \label{eq:fullprefactor}
\end{align}
The local saddle integral supplies $1/p$, while $\alpha$ records the normalization of the complete measure.

This separation also explains why the constant in the normalized moment is more robust than the logarithm of the complete integral.
The former is an expectation value of the scaled local potential.
The latter counts the shrinking volume of configuration space as well as any normalization already present in $\dd\nu_g$.
A regular change of coordinates changes $\rho_0$ and $\widehat a_p$ in compensating fashion, leaving both the integral and its normalized moment invariant.

Let us consider a non-Gaussian example.
An ordinary critical point is the codimension-two endpoint obtained by tuning the bath temperature and one additional control variable while the remaining boundary data are fixed.
Let $x$ be a dimensionless coordinate along its single thermodynamic soft direction, with $S'(x_c)$ finite and nonzero.
The restricted first law gives
\begin{equation}
 \frac{\dd\Fgen}{\dd x}=[\TH(x)-\TE]S'(x).
 \label{eq:xfirstderivative}
\end{equation}
Differentiating once gives $\Fgen''=\TH'S'+(\TH-\TE)S''$.
At the critical point $\TE=T_c=\TH(x_c)$ and $\TH'(x_c)=\TH''(x_c)=0$.
The first three derivatives of $\Fgen$ vanish, while only the term proportional to $\TH'''S'$ survives in the fourth derivative, giving
\begin{equation}
 \Fgen''''(x_c)=\TH'''(x_c)S'(x_c)>0.
 \label{eq:xcriticalderivatives}
\end{equation}
The critical saddle therefore has $p=4$.
For a smooth action and measure, the proposition gives
\begin{equation}
 \Fav_c-F_c=\frac{T_c}{4}+\mathcal O(T_cg^{1/2}),
 \qquad Z_{\mathrm{loc}}\propto g^{1/4}.
 \label{eq:quarter}
\end{equation}
Every noncritical direction retains its Gaussian contribution $\TE/2$.
Equation~\eqref{eq:quarter} replaces only the contribution of the critical mode.

\subsection{Multi-dimensional saddle-order universality}
\label{sec:multidimensional}

The one-dimensional proposition can be extended without assuming that the leading collective directions decouple.
Let
\begin{equation}
 \bm q=(q^1,\ldots,q^n),\qquad
 \bm\delta=\bm q-\bm q_0,
 \label{eq:multiq}
\end{equation}
where $\bm q_0$ is a unique stable interior minimum of the reduced action.
The measure generalizing Eq.~\eqref{eq:measurefactor} is
\begin{equation}
 \dd\nu_g(\bm q)=\mathcal N(g)\rho(\bm q,g)\dd^n q,
 \qquad
 \rho(\bm q_0,g)\rightarrow \rho_0>0,
 \label{eq:multimeasure}
\end{equation}
and the source-dependent partition function retains the definition in Eq.~\eqref{eq:sourceZ},
\begin{equation}
 Z(s)=\int_{\mathcal D}\dd\nu_g(\bm q)e^{-s\IE(\bm q)}.
 \label{eq:multiZ}
\end{equation}
Assume that, in a $g$-independent local coordinate system,
\begin{equation}
 \IE(\bm q_0+\bm\delta)
 =I_0+\frac{1}{g}V_0(\bm\delta)+\cdots,
 \label{eq:multiaction}
\end{equation}
where $V_0$ is positive away from the origin on the chosen real contour and is weighted homogeneous, so that
\begin{equation}
 V_0\!\left(
 \lambda^{w_1}\delta^1,\ldots,\lambda^{w_n}\delta^n
 \right)
 =\lambda V_0(\bm\delta),
 \qquad w_i>0.
 \label{eq:weightedhomogeneous}
\end{equation}
Define the total scaling weight
\begin{equation}
 W\equiv\sum_{i=1}^{n}w_i,
 \label{eq:Wdef}
\end{equation}
and suppose that
\begin{equation}
 C_V\equiv\int_{\mathbb R^n}\dd^n x\,e^{-V_0(\bm x)}
 \label{eq:CVdef}
\end{equation}
is finite.
The omitted terms in Eq.~\eqref{eq:multiaction} are assumed
to be strictly subleading under the weighted rescaling below.
We further assume that the density approaches $\rho_0$
uniformly in the rescaled saddle region, with sufficient
control of the tails to interchange the asymptotic expansion
with the local integration. These estimates are taken to hold
uniformly for $s$ in a neighborhood of unity and after one
derivative with respect to $s$.

\begin{proposition}
\label{prop:multisaddle}
Under the assumptions above, a regular weighted-homogeneous interior saddle contributes
\begin{equation}
 \begin{aligned}
 \Fav-\Fcl&=\TE W+o(\TE)\\
 &=\TE\sum_{i=1}^{n}w_i+o(\TE),
 \qquad
 \Fcl=\frac{I_0}{\beta}.
 \end{aligned}
 \label{eq:multimainlaw}
\end{equation}
while its local integral scales as
\begin{equation}
 Z_{\rm loc}(s)
 =e^{-sI_0}\mathcal N(g)\rho_0 C_V\,
 g^W s^{-W}\,[1+o(1)].
 \label{eq:multiprefactor}
\end{equation}
where the little-$o$ terms refer to the limit $g\to0$ at fixed $T_
E$. 
The statement is local. Other saddles must remain exponentially separated. The density must remain smooth and nonzero near the saddle, which must stay a finite distance from the boundary of the physical integration domain.
\end{proposition}

\begin{propositionproof}
Use the weighted rescaling
\begin{equation}
 \delta^i=\left(\frac{g}{s}\right)^{w_i}x^i.
 \label{eq:weightedrescaling}
\end{equation}
Its Jacobian is
\begin{equation}
 \dd^n\delta
 =\left(\frac{g}{s}\right)^W\dd^n x.
 \label{eq:weightedjacobian}
\end{equation}
By Eq.~\eqref{eq:weightedhomogeneous},
\begin{equation}
 \frac{s}{g}V_0(\bm\delta)=V_0(\bm x).
 \label{eq:weightedactionscale}
\end{equation}
Keeping the leading value $\rho_0$ of the local density therefore gives Eq.~\eqref{eq:multiprefactor}.
Taking its logarithm,
\begin{equation}
 \ln Z_{\rm loc}(s)
 =-sI_0-W\ln s+W\ln g
 +\ln[\mathcal N(g)\rho_0C_V]+o(1),
 \label{eq:multilogZ}
\end{equation}
and substituting this result into Eq.~\eqref{eq:sourceFav} immediately yields Eq.~\eqref{eq:multimainlaw}.
\end{propositionproof}

There is an equivalent virial proof that makes the relation to Eq.~\eqref{eq:virial} explicit.
Weighted Euler homogeneity gives
\begin{equation}
 \sum_{i=1}^{n}w_i x^i\frac{\partial V_0}{\partial x^i}=V_0.
 \label{eq:weightedeuler}
\end{equation}
Assuming the boundary contribution vanishes,
\begin{align}
 0
 &=\sum_{i=1}^{n}\int\dd^n x\,
 \frac{\partial}{\partial x^i}
 \left(w_i x^i e^{-V_0}\right) \notag\\
 &=W\int\dd^n x\,e^{-V_0}
 -\int\dd^n x\,V_0e^{-V_0}.
 \label{eq:multivirialproof}
\end{align}
Consequently
\begin{equation}
 \langle V_0\rangle_0=W,
 \label{eq:multivirial}
\end{equation}
which is the weighted multivariable form of generalized equipartition \cite{Tolman1918}.
For a single monomial $V_0\propto x^p$, Eq.~\eqref{eq:weightedhomogeneous} gives $w=1/p$, and Proposition~\ref{prop:multisaddle} reduces exactly to Proposition~\ref{prop:saddleorder}.

A generic analytic potential need not be weighted homogeneous with respect to the naive scaling of its individual monomials.
Appendix~\ref{app:general-polynomial} shows how additional polynomial couplings can either remain subleading, enter the leading weighted potential, or reorganize the dominant saddle scaling altogether.

\subsection{Decoupled and coupled examples}
\label{subsec:coupledexamples}

For independent stable monomials
\begin{equation}
 V_0(\bm x)=\sum_{i=1}^{n}a_i(x^i)^{p_i},
 \qquad
 a_i>0,\quad p_i\ {\rm even},
 \label{eq:independentmonomials}
\end{equation}
one has $w_i=1/p_i$ and therefore
\begin{equation}
 \frac{\Fav-\Fcl}{\TE}
 =\sum_{i=1}^{n}\frac{1}{p_i}.
 \label{eq:independentlaw}
\end{equation}
Thus two Gaussian modes give $\Fav-\Fcl=\TE$, one Gaussian plus one quartic mode gives $3\TE/4$, and two quartic modes give $\TE/2$.
The last example shows that the coefficient $\TE/2$ does not, by itself, imply that the underlying saddle is Gaussian once more than one soft direction is present.

A useful nonfactorized example contains one quartic and one Gaussian scaling direction,
\begin{equation}
 V_0(x,y)=\frac{a}{2}y^2+b x^2y+c x^4.
 \label{eq:gaussquarticcoupled}
\end{equation}
It is weighted homogeneous with
\begin{equation}
 w_x=\frac14,\qquad w_y=\frac12,
 \label{eq:gaussquarticweights}
\end{equation}
since all three terms have weighted degree one.
Proposition~\ref{prop:multisaddle} therefore predicts
\begin{equation}
 \Fav-\Fcl=\frac{3\TE}{4}.
 \label{eq:threequarterprediction}
\end{equation}
This prediction can be checked directly.
Completing the square gives
\begin{equation}
 V_0
 =\frac{a}{2}\left(y+\frac{b}{a}x^2\right)^2
 +c_{\rm eff}x^4,
 \qquad
 c_{\rm eff}=c-\frac{b^2}{2a}.
 \label{eq:completecoupledsquare}
\end{equation}
For
\begin{equation}
 a>0,\qquad c_{\rm eff}>0,
 \label{eq:coupledstability}
\end{equation}
the shift $Y=y+(b/a)x^2$ has unit Jacobian. In these variables the local source integral factorizes as
\begin{align}
 Z_{\rm loc}(s)
 &\propto
 \int_{-\infty}^{\infty}\dd Y\,
 e^{-saY^2/(2g)}
 \int_{-\infty}^{\infty}\dd x\,
 e^{-s c_{\rm eff}x^4/g} \notag\\
 &=
 \sqrt{\frac{2\pi g}{sa}}\,
 \frac{\Gamma(1/4)}{2}
 \left(\frac{g}{s c_{\rm eff}}\right)^{1/4}.
 \label{eq:coupledexplicitZ}
\end{align}
Hence $Z_{\rm loc}\propto g^{3/4}s^{-3/4}$ and Eq.~\eqref{eq:threequarterprediction} follows from the source derivative.

As a genuinely quartic two-mode example, consider
\begin{equation}
 V_0(x,y)=a x^4+b x^2y^2+c y^4.
 \label{eq:twoquarticpotential}
\end{equation}
Both variables have weight $1/4$, so
\begin{equation}
 Z_{\rm loc}\propto g^{1/2},
 \qquad
 \Fav-\Fcl=\frac{\TE}{2}.
 \label{eq:twoquarticresult}
\end{equation}
A sufficient stability condition on the real plane is
\begin{equation}
 a>0,\qquad c>0,\qquad b>-2\sqrt{ac}.
 \label{eq:twoquarticstability}
\end{equation}
Equation~\eqref{eq:twoquarticresult} provides a simple counterexample to identifying the numerical value $\TE/2$ uniquely with one Gaussian direction.

More generally, if $n_{\rm G}$ locally massive directions remain Gaussian while the soft sector is described by weighted coordinates with weights $w_a$, the leading counting law is
\begin{equation}
 \frac{\Fav-\Fcl}{\TE}
 =\frac{n_{\rm G}}{2}+\sum_a w_a,
 \qquad
 Z_{\rm loc}\propto
 g^{\,n_{\rm G}/2+\sum_a w_a}.
 \label{eq:massivesoftmaster}
\end{equation}
For one quartic soft direction, Eq.~\eqref{eq:massivesoftmaster} reduces to the $1/4+n_{\rm G}/2$ counting quoted below in the discussion, but the weighted formulation also covers regular couplings within the critical sector.
These examples illustrate the mathematical content of the multidimensional saddle-order proposition before applying it to concrete black hole thermodynamic systems.

\section{RN-AdS realization}
\label{sec:rn}

\subsection{Thermodynamics and critical point}

We now ask whether the weighted saddle structure appears in actual black hole thermodynamic landscapes.
The four-dimensional RN-AdS free-energy landscape at fixed $(Q,L)$ is obtained by evaluating the Euclidean action on the off-shell family of geometries parametrized by the horizon radius $r$. The boundary period $\beta=\TE^{-1}$ is held fixed, while regularity at the horizon is not imposed away from a stationary configuration. The resulting geometries therefore generally contain a conical defect at the horizon. The reduced Euclidean action is
\begin{equation}
\IE(r;\TE)
=\frac{\beta}{2\GN}
\left(r+\frac{r^3}{L^2}+\frac{Q^2}{r}\right)
-\frac{\pi r^2}{\GN},
\label{eq:rnaction}
\end{equation}
which defines the generalized free energy through $\Fgen=\TE\IE$,
\begin{equation}
\Fgen(r;\TE)=\frac{1}{2\GN}
\left(r+\frac{r^3}{L^2}+\frac{Q^2}{r}-2\pi\TE r^2\right).
\label{eq:rnfree}
\end{equation}

At a stationary point, the conical defect disappears and the geometry becomes a smooth RN-AdS black hole. Indeed, differentiating Eq.~\eqref{eq:rnfree} with respect to $r$ at fixed $(Q,\TE,L)$ gives a condition equivalent to $\TH(r)=\TE$, 
%
where 
\begin{equation}
	\TH=\frac{1}{4\pi r}\left(1+\frac{3r^2}{L^2}-\frac{Q^2}{r^2}\right)\,.
\end{equation}
The critical data are given in Refs.~\cite{Chamblin1999,CaldarelliCognolaKlemm2000,KubiznakMann2012, KubiznakMannTeo2017} as
\begin{equation}
 r_c=\frac{L}{\sqrt6}=\sqrt6\,Q_c,\qquad
 Q_c=\frac{L}{6},\qquad
 T_c=\frac{\sqrt6}{3\pi L}
 \label{eq:rncritical}
\end{equation}
At these values
\begin{align}
 \Fgen'(r_c)&=\Fgen''(r_c)=\Fgen'''(r_c)=0, \notag\\
 \Fgen''''(r_c)&=\frac{12Q_c^2}{\GN r_c^5}>0,
 \label{eq:rnderivatives}
\end{align}
which verifies the stable quartic saddle.


To evaluate the critical integral, introduce the dimensionless variables
\begin{equation}
 g=\frac{\GN}{L^2},\qquad z=\frac rL,
 \qquad q=\frac QL,\qquad \tau=L\TE.
 \label{eq:rndimensionless}
\end{equation}
The mass, entropy, and generalized free energy have the dimensionless forms
\begin{align}
 m(z,q)&=\frac{\GN M}{L}
 =\frac12\left(z+z^3+\frac{q^2}{z}\right), \notag\\
 \sigma(z)&=gS=\pi z^2, \notag\\
 f(z;\tau,q)&=\frac{\GN\Fgen}{L}=m(z,q)-\pi\tau z^2.
 \label{eq:rndimensionlessthermo}
\end{align}
The reduced Euclidean action is
\begin{equation}
 \IE(z;\tau,q)=\frac{\mathcal I(z;\tau,q)}{g},
\end{equation}
with
\begin{equation}
 \mathcal I=\frac{z+z^3+q^2/z-2\pi\tau z^2}{2\tau}.
 \label{eq:rndimensionlessaction}
\end{equation}
The common factor $g^{-1}=L^2/\GN$ is the large parameter of the saddle expansion.
It multiplies the entire dimensionless action rather than an individual term, so the coefficients of the Taylor series of $\mathcal I$ remain finite at fixed $(z,q,\tau)$.
This is the scaling assumed in Proposition~\ref{prop:saddleorder}.

All quantities inside $\mathcal I$ remain fixed as $g\to0$.
The unrescaled Bekenstein-Hawking entropy is $S=\pi z^2/g$.
Using $S$ as the asymptotic coordinate would therefore move powers of $g$ between the quartic coefficient and the measure Jacobian without changing the physical local integral.

Within the one-coordinate model, choose the dimensionless energy measure
\begin{equation}
 \dd m=\rho_E(z,q)\dd z,
 \qquad
 \rho_E(z,q)=\frac12\left(1+3z^2-\frac{q^2}{z^2}\right).
 \label{eq:rnmeasure}
\end{equation}
It is obtained by changing variables from the physical mass along the off-shell family.
Since $m=\GN M/L$, one has $\rho_E=(\dd m/\dd z)_q$ along the fixed-charge off-shell family.
This choice weights equal intervals of dimensionless energy equally and supplies a direct physical meaning for the reduced probability distribution.
It is a nonunique reduced prescription rather than a measure derived from the complete gauge-fixed gravitational path integral; any alternative density that is smooth and nonzero at the critical saddle leaves the normalized $T_c/4$ contribution unchanged.
Near a regular stationary black hole it is related to the entropy derivative by the first law.

It is positive on the nonextremal branch $z>z_e(q)$ and vanishes at the extremal endpoint.
Indeed, the dimensionless Hawking temperature $\tau_H=L\TH$ obeys
\begin{equation}
 \rho_E(z,q)=2\pi z\tau_H(z,q),
 \qquad
 z_e^2(q)=\frac{-1+\sqrt{1+12q^2}}{6}.
 \label{eq:rnmeasuretemperature}
\end{equation}
The critical saddle lies strictly inside this branch because $z_e(q_c)<z_c$.
The finite separation leaves the local critical expansion unchanged, while the positivity of the density near $z_c$ ensures that the energy coordinate is regular.
The original energy measure satisfies
\begin{equation}
 \dd M=\frac{L}{\GN}\dd m=\frac{1}{gL}\dd m.
 \label{eq:rawenergymeasure}
\end{equation}
The factor $1/(gL)$ is independent of $z$.
It cancels from normalized expectation values but would add a measure-dependent power of $g$ to an unnormalized partition function.
We use $\dd m$ below, so the local density is dimensionless and independent of $g$.


At the critical point
\begin{equation}
 z_c=\frac1{\sqrt6},\qquad q_c=\frac16,
 \qquad \tau_c=\frac{\sqrt6}{3\pi},
 \qquad \mathcal I_c=\frac\pi6.
 \label{eq:rncriticaldimensionless}
\end{equation}
Here $\mathcal I_c$ is the dimensionless critical action, while $I_{\mathrm E,c}\equiv\mathcal I_c/g=\beta_cF_c$, with $\beta_c=T_c^{-1}$ and $F_c$ the on-shell generalized free energy of the critical black hole.
Writing $\delta=z-z_c$, direct expansion gives
\begin{equation}
 \mathcal I-\mathcal I_c
 =\frac{3\pi}{2}\delta^4
 -\frac{3\pi\sqrt6}{2}\delta^5
 +9\pi\delta^6+\mathcal O(\delta^7),
 \label{eq:rncriticalexpansion}
\end{equation}
while
\begin{equation}
 \rho_E(z_c,q_c)=\frac23.
 \label{eq:rncriticalmeasure}
\end{equation}
The absence of quadratic and cubic terms is the local form of the two inflection conditions.
The positive coefficient $3\pi/2$ identifies $\widehat a_4$ in the proposition.
Balancing $(3\pi/2)\delta^4/g$ against unity gives the width without an additional thermodynamic assumption.

The critical width in the $g$-independent coordinate is
\begin{equation}
 \Delta z=\left(\frac{2g}{3\pi}\right)^{1/4}.
 \label{eq:rnwidth}
\end{equation}
The physical lower endpoint remains a finite distance from $z_c$.
Its effect is exponentially small, and the leading local integral can be extended to the real line.
With $C_4=\Gamma(1/4)/2$ and the source of Eq.~\eqref{eq:sourceZ}, one obtains
\begin{equation}
 Z_c(s)=e^{-s\pi/(6g)}\frac23 C_4
 \left(\frac{2g}{3\pi s}\right)^{1/4}
 \left[1+\mathcal O(g^{1/2})\right].
 \label{eq:rncriticalZ}
\end{equation}
The quintic action term and the first derivative of $\rho_E$ produce odd integrands at order $g^{1/4}$.
Their integrals vanish.
The sextic term and products of the odd corrections enter at order $g^{1/2}$.
Applying the source derivative in Eq.~\eqref{eq:sourceFav} removes the classical term $I_{\mathrm E,c}=\pi/(6g)$ when $F_c$ is subtracted and differentiates the factor $s^{-1/4}$.
It yields
\begin{equation}
 \frac{\Fav-F_c}{T_c}=\frac14+\mathcal O(g^{1/2}).
 \label{eq:rnquarteranalytic}
\end{equation}
This verifies the soft-mode prediction with the chosen reduced energy measure.

The calculation also shows where the leading error originates.
Rescaling $\delta=\Delta z\,x$ makes the quintic term proportional to $g^{1/4}x^5$ and the linear variation of the density proportional to $g^{1/4}x$.
The surviving terms at the next order come from the sextic action, the square of the quintic perturbation, and its product with the linear density variation.

\subsection{Control fields}

The RN-AdS action also fixes the control fields of the cusp.
Let $\Delta\tau=\tau-\tau_c$ and $\Delta q=q-q_c$.
Expanding about $z_c$ and translating the local coordinate to remove the cubic term gives
\begin{equation}
 \mathcal I=\mathcal I_0(\tau,q)-wy+uy^2
 +\frac{3\pi}{2}y^4
 +\mathcal O(y^5,\Delta y^3,\Delta^2y).
 \label{eq:rnnormalform}
\end{equation}
The quantity $\mathcal I_0(\tau,q)$ is the control-dependent constant term of the dimensionless action and equals $\mathcal I_c$ only at $(\tau_c,q_c)$.
The symbol $\Delta$ in the remainder denotes the common order of the two control displacements $\Delta\tau$ and $\Delta q$.
The field $w$ tilts the potential and breaks the symmetry between the two sides of the critical radius.
The field $u$ changes the local curvature.
Positive $u$ gives one minimum near the origin, while negative $u$ permits the pair of minima associated with the coexistence region.

Here
\begin{align}
 u&=3\pi\Delta q-\frac{\sqrt6\pi^2}{2}\Delta\tau,
 \notag\\
 w&=\pi^2\Delta\tau+\frac{\sqrt6\pi}{2}\Delta q.
 \label{eq:rnfields}
\end{align}
Before the translation, the terms linear and quadratic in $\delta=z-z_c$ have coefficients
\begin{equation}
 -\left(\pi^2\Delta\tau+\frac{\sqrt6\pi}{2}\Delta q\right)
 \quad\hbox{and}\quad
 3\pi\Delta q-\frac{\sqrt6\pi^2}{2}\Delta\tau,
 \label{eq:rnfieldcoefficients}
\end{equation}
respectively.
The control-dependent cubic coefficient is removed by a shift of order $\Delta$.
Such a shift changes the displayed linear map only at second order in the controls, which is already contained in the remainder of Eq.~\eqref{eq:rnnormalform}.
The Jacobian $\partial(u,w)/\partial(\tau,q)=-9\pi^3/2$ is nonzero, so $u$ and $w$ are independent local fields.
The control-dependent constant $\mathcal I_0(\tau,q)$ must be retained when constructing the partition function.
At fixed $(\tau,q)$, $\Fcl$ denotes the generalized free energy evaluated at the global minimum $z_*(\tau,q)$; hence $\Fcl=F_c$ only at the critical point.

The nonzero Jacobian is the local statement that temperature and charge can independently tune curvature and tilt.
Retaining $\mathcal I_0$ is equally important for a thermodynamic comparison between control points, even though it cancels when the fluctuation contribution is measured relative to the classical minimum at the same $(\tau,q)$.

The zero-field tangent direction obeys
\begin{equation}
 \Delta\tau=-\frac{\sqrt6}{2\pi}\Delta q,
 \qquad u=\frac{9\pi}{2}\Delta q.
 \label{eq:rnzerofieldpath}
\end{equation}
The side with $\Delta q\geq0$ has $u\geq0$ and a single local minimum.
It is the RN-AdS realization of the one-well crossover studied below.

\subsection{Uniform critical crossover}
\label{sec:crossover}

Exactly at criticality the quartic term is sufficient, but a generic neighborhood has two relevant control fields.
To describe this neighborhood, write the normal form of Eq.~\eqref{eq:rnnormalform} with explicit coefficients
\begin{equation}
 \IE-\frac{\mathcal I_0(\tau,q)}{g}
 =\frac1g\left(c_4y^4+c_2uy^2-c_1wy\right)+\cdots .
 \label{eq:cusp}
\end{equation}
The coefficients $c_i$ are positive.
Here $u=w=0$ is the critical point.
Introducing
\begin{equation}
 x=\left(\frac{c_4}{g}\right)^{1/4}y,
 \quad \lambda=\frac{c_2u}{\sqrt{c_4g}},
 \quad h=\frac{c_1w}{c_4^{1/4}g^{3/4}},
 \label{eq:scaledfields}
\end{equation}
gives the universal potential and integral
\begin{equation}
 \mathcal V=x^4+\lambda x^2-hx,\qquad
 \mathcal Z(\lambda,h)=\int_{-\infty}^{\infty}\dd x\,e^{-\mathcal V}.
 \label{eq:scaledintegral}
\end{equation}
The powers of $g$ follow by balancing each perturbation against the quartic weight.
Since $y=\mathcal O(g^{1/4})$, the quadratic perturbation is of order unity when $u=\mathcal O(g^{1/2})$, and the linear perturbation is of order unity when $w=\mathcal O(g^{3/4})$.
Holding $\lambda$ and $h$ fixed as $g\to0$ keeps all three terms in the scaled potential finite.
After factoring out $\mathcal N(g)$, define $\mu_0(\tau,q)\equiv\lim_{g\to0}\rho_y(0;\tau,q,g)$ with $\rho_y=\rho_z|\dd z/\dd y|$; for the RN-AdS energy measure, $\rho_z=\rho_E$ and $\mu_c\equiv\mu_0(\tau_c,q_c)=2/3$.

In particular, the local partition function becomes
\begin{equation}
 \begin{aligned}
 Z_{\mathrm{loc}}={}&\exp\!\left[-\frac{\mathcal I_0(\tau,q)}{g}\right]
 \mathcal N(g)\mu_0(\tau,q) \\
 &\times\left(\frac{g}{c_4}\right)^{1/4}
 \mathcal Z(\lambda,h)\left[1+\mathcal O(g^{1/4})\right].
 \end{aligned}
 \label{eq:uniformZ}
\end{equation}
Equation~\eqref{eq:uniformZ} is the uniform replacement for the singular Gaussian determinant.
Canonical integrals play the same role in classical uniform asymptotics for coalescing saddles \cite{ChesterFriedmanUrsell1957}.
The two-field quartic form is the real-weight counterpart of the cusp integral introduced by Pearcey \cite{Pearcey1946}.
Equivalently, for $\Geff\equiv-\ln Z_{\mathrm{loc}}$,
\begin{equation}
 \begin{aligned}
 \Geff-\frac{\mathcal I_0(\tau,q)}{g}
 ={}&\frac14\ln\!\left(\frac{c_4}{g}\right)
 -\ln\mathcal Z(\lambda,h) \\
 &-\ln\!\left[\mathcal N(g)\mu_0(\tau,q)\right]
 +\mathcal O(g^{1/4}),
 \end{aligned}
 \label{eq:uniformGamma}
\end{equation}
Without a derivation of $\mathcal N(g)$ from the complete gauge-fixed path integral, Eq.~\eqref{eq:uniformGamma} does not determine the logarithmic coefficient of the complete gravitational partition function.
The coefficient $1/4$ replaces the Gaussian $1/2$ in the soft part of the logarithmic prefactor.

For RN-AdS, $c_4=3\pi/2$ and the leading normal form has $c_2=c_1=1$. 
Consequently the fields in Eq.~\eqref{eq:rnfields} determine the scaling variables without a numerical fit.
The line $w=0$ selects $h=0$, and its $u\geq0$ side remains in the one-minimum sector required for a direct normalized average around one saddle.

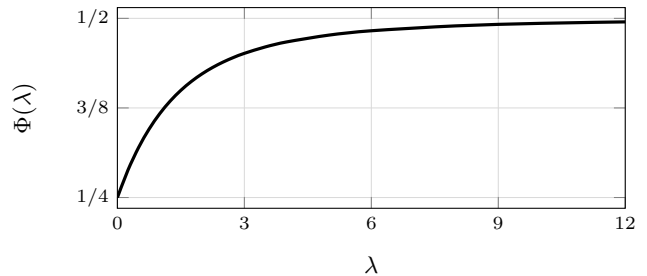
\begin{figure}[tbp]
\centering
\begin{tikzpicture}
\begin{axis}[
  width=0.96\columnwidth,
  height=0.49\columnwidth,
  xmin=0,xmax=12,
  ymin=0.235,ymax=0.515,
  xlabel={$\lambda$},
  ylabel={$\Phi(\lambda)$},
  xtick={0,3,6,9,12},
  ytick={0.25,0.375,0.5},
  yticklabels={$1/4$,$3/8$,$1/2$},
  tick label style={font=\scriptsize},
  label style={font=\small},
  axis line style={black},
  grid=major,
  grid style={line width=0.2pt,draw=gray!30},
  every axis plot/.append style={line join=round}
]
\addplot[black,very thick,smooth] coordinates {
  (0,0.25000000)
  (0.25,0.28829292)
  (0.5,0.31971100)
  (0.75,0.34558351)
  (1,0.36697998)
  (1.25,0.38475730)
  (1.5,0.39960053)
  (1.75,0.41205708)
  (2,0.42256475)
  (2.25,0.43147429)
  (2.5,0.43906747)
  (2.75,0.44557135)
  (3,0.45116955)
  (3.5,0.46021786)
  (4,0.46710676)
  (5,0.47663418)
  (6,0.48268044)
  (8,0.48951871)
  (10,0.49303596)
  (12,0.49505865)
};
\end{axis}
\end{tikzpicture}
\caption{Zero-field one-well scaling function. The curve is exact for the quartic normal-form integral and gives the leading RN-AdS crossover in the double-scaling limit at fixed $\lambda$ and $h$. The quartic value $1/4$ occurs at $\lambda=0$, and the Gaussian value $1/2$ is recovered as $\lambda\to\infty$. For RN-AdS the scaling variable is $\lambda=u/\sqrt{(3\pi/2)g}$ with $u$ given in Eq.~\eqref{eq:rnfields}.}
\label{fig:crossover}
\end{figure}

If $x_*$ is the unique global minimum of $\mathcal V$, the normalized generalized free energy contribution is
\begin{equation}
 \frac{\Fav-\Fcl}{\TE}
 =\langle\mathcal V\rangle-\mathcal V(x_*)+\mathcal O(g^{1/4}).
 \label{eq:scalingfunction2d}
\end{equation}
Along the zero-field one-well line $h=0$, $\lambda\geq0$, one has $x_*=0$ and
\begin{align}
 \mathcal Z(\lambda)
 &=\int_{-\infty}^{\infty}\dd x\,e^{-x^4-\lambda x^2} \notag\\
 &=\frac{\sqrt\lambda}{2}e^{\lambda^2/8}
 K_{1/4}\!\left(\frac{\lambda^2}{8}\right),
 \qquad \lambda>0.
 \label{eq:bessel}
\end{align}
Integration by parts yields an exact crossover function for the quartic normal-form integral,
\begin{align}
 \Phi(\lambda)&\equiv\frac{\Fav-\Fcl}{\TE}
 =\frac14-\frac{\lambda}{2}\frac{\dd}{\dd\lambda}\ln\mathcal Z(\lambda),
 \notag\\
 \Phi(0)&=\frac14,\qquad \Phi(\infty)=\frac12.
 \label{eq:phi}
\end{align}
For the complete RN-AdS reduced integral, Eq.~\eqref{eq:phi} is the leading double-scaling result obtained as $g\to0$ with $\lambda$ and $h$ fixed.
Corrections arise from the omitted terms in the action and from the measure; the physical endpoint remains outside the scaling region.
The identity behind this expression is $1=4\langle x^4\rangle+2\lambda\langle x^2\rangle$.
Since $-\dd\ln\mathcal Z/\dd\lambda=\langle x^2\rangle$, it converts the mean potential into the derivative shown in Eq.~\eqref{eq:phi}.
At $\lambda=0$ only the quartic term remains and the proposition gives $1/4$.
For large $\lambda$, the typical coordinate is $x=\mathcal O(\lambda^{-1/2})$ and the quartic term becomes subleading, leaving the Gaussian value $1/2$.
Figure~\ref{fig:crossover} shows this crossover for the normal-form integral.
Thus $u=0$ followed by $g\to0$ gives $1/4$, whereas fixed $u>0$ followed by $g\to0$ sends $\lambda\to\infty$ and restores $1/2$.
The two values therefore describe different orders of the critical and semiclassical limits.

The same normal form also locates the coexistence sector.
When $\lambda<0$ at $h=0$, it has two equal minima, while a nonzero $h$ selects one of them.
The RN-AdS first-order small-large black hole transition lies in this multiple-minimum region.
Its normalized average includes both saddle weights and depends on the contour through the complete reduced landscape.
The analytic function in Eq.~\eqref{eq:phi} instead describes the one-well path in Eq.~\eqref{eq:rnzerofieldpath}, for which a single stable saddle controls the integral.

\section{Kerr-AdS realization}
\label{sec:kerr}

Kerr-AdS provides a rotating realization of the saddle-order proposition. Its equilibrium critical point at fixed angular momentum is well established~\cite{WeiLiu2021,CaldarelliCognolaKlemm2000}, while its ensemble-averaged thermodynamics in the Gaussian regime was studied in Ref.~\cite{ChengPanXuYang2025}. Critical slowing down in Kerr-AdS has also been investigated through stochastic evolution of the entropy on the free energy landscape~\cite{AwalPhukon2026}. 
Here the purpose is not to repeat the RN-AdS saddle analysis, but to show that the Kerr-AdS critical point admits the same quartic structure with a regular reduced measure and an analytic map from the physical controls to the universal cusp variables.

At fixed angular momentum $J$ and AdS radius $L$, the canonical generalized free energy is
\begin{equation}
 \Fgen(S;\TE,J)=M(S,J,L)-\TE S .
 \label{eq:kerrFgen}
\end{equation}
The first law implies, at fixed $(J,L)$,
\begin{equation}
 \left(\frac{\partial \Fgen}{\partial S}\right)_{J,L}
 =\TH-\TE .
 \label{eq:kerrstationaryphysical}
\end{equation}
The stationary members of the off-shell entropy family are therefore precisely the smooth Kerr-AdS black holes satisfying $\TH=\TE$.

We use the same semiclassical parameter as in the RN-AdS analysis,
\begin{equation}
 g=\frac{\GN}{L^2},
\end{equation}
and define
\begin{equation}
 \sigma=gS,
 \qquad
 j=gJ,
 \qquad
 m=\frac{\GN M}{L},
 \qquad
 \tau=L\TE .
 \label{eq:kerrdimensionless}
\end{equation}
The semiclassical limit is taken with $(\tau,j)$ fixed.
The dimensionless generalized free energy and reduced action are
\begin{equation}
 \begin{aligned}
 f(\sigma;\tau,j)
 &=\frac{\GN\Fgen}{L}
 =m(\sigma,j)-\tau\sigma,\\
 \IE
 &=\frac{\mathcal I_K}{g},
 \qquad
 \mathcal I_K(\sigma;\tau,j)
 =\frac{m(\sigma,j)}{\tau}-\sigma .
 \end{aligned}
 \label{eq:kerractiondimensionless}
\end{equation}
Thus the entire semiclassical dependence again appears through the overall factor $g^{-1}$.

It is useful to introduce the $g$-independent coordinate
\begin{equation}
 \chi=\frac{\sigma}{\pi}.
\end{equation}
The Kerr-AdS mass then takes the compact form
\begin{equation}
 m(\chi,j)=\frac12\sqrt{\mathcal R(\chi,j)},
\end{equation}
with
\begin{equation}
 \mathcal R(\chi,j)
 =
 \chi^3+2\chi^2+\chi+4j^2+\frac{4j^2}{\chi}.
 \label{eq:kerrRdef}
\end{equation}
Since $\partial_\sigma=\pi^{-1}\partial_\chi$, the dimensionless Hawking temperature becomes
\begin{equation}
 \tau_H(\chi,j)
 =
 \frac{
 3\chi^2+4\chi+1-4j^2/\chi^2
 }{
 4\pi\sqrt{\mathcal R}
 } .
 \label{eq:kerrtauH}
\end{equation}
Consequently,
\begin{equation}
 \frac{\partial\mathcal I_K}{\partial\sigma}
 =
 \frac{\tau_H-\tau}{\tau},
\end{equation}
which makes the relation between the reduced saddle and the equilibrium black hole explicit.

At fixed $j$, the ordinary critical endpoint obeys
\begin{equation}
 m_{\sigma\sigma}=0,
 \qquad
 m_{\sigma\sigma\sigma}=0 .
 \label{eq:kerrcriticalconditions}
\end{equation}
For $m=\sqrt{\mathcal R}/2$, the first condition gives
\begin{equation}
 2\mathcal R\mathcal R''-(\mathcal R')^2=0,
\end{equation}
while, after this relation is imposed, the second condition reduces to
\begin{equation}
 \mathcal R'''_c=0 .
\end{equation}
These two equations give
\begin{equation}
 j_c=\frac{\chi_c^2}{2},
 \qquad
 8\chi_c^3+10\chi_c^2+2\chi_c-1=0 .
 \label{eq:kerrcubic}
\end{equation}
The cubic is strictly increasing for $\chi>0$ and therefore possesses a unique positive root.
The critical entropy and temperature are then
\begin{equation}
 \sigma_c=\pi\chi_c,
 \qquad
 \tau_c=
 \frac{B_c}{4\pi\sqrt{\mathcal R_c}},
 \qquad
 B_c=2\chi_c^2+4\chi_c+1,
 \label{eq:kerrcriticaldata}
\end{equation}
where $\mathcal R_c=\mathcal R(\chi_c,j_c)$.

At $(\tau_c,j_c)$, the first three derivatives of $\mathcal I_K$ vanish.
The first stabilizing term is quartic, with
\begin{equation}
 \widehat a_{4,K}
 \equiv
 \frac{m_{\sigma\sigma\sigma\sigma,c}}{24\tau_c}
 =
 \frac{1}{\pi^3\chi_c B_c}
 >0 .
 \label{eq:kerra4}
\end{equation}
The Kerr-AdS critical point is therefore a stable quartic saddle.
Its characteristic width is
\begin{equation}
 \Delta\sigma
 =
 \left(
 \frac{g}{\widehat a_{4,K}}
 \right)^{1/4}
 =
 \left(\pi^3\chi_cB_c\right)^{1/4}g^{1/4}.
 \label{eq:kerrwidth}
\end{equation}
Rotation changes the black-hole-dependent coefficient, but not the $g^{1/4}$ critical scaling.

For the reduced ensemble we again use the dimensionless energy measure along the off-shell family,
\begin{equation}
 \dd m=\rho_{E,K}(\sigma,j)\,\dd\sigma .
 \label{eq:kerrmeasure}
\end{equation}
The first law gives
\begin{equation}
 \rho_{E,K}=m_\sigma=\tau_H,
\end{equation}
so that
\begin{equation}
 \rho_{E,K}(\sigma_c,j_c)=\tau_c>0 .
\end{equation}
The energy coordinate is therefore regular at the critical saddle.

The critical point also lies strictly inside the nonextremal integration domain.
At fixed $j_c$, the extremal endpoint is determined by $\tau_H=0$, or
\begin{equation}
 3\chi_e^4+4\chi_e^3+\chi_e^2-\chi_c^4=0 .
 \label{eq:kerrextremal}
\end{equation}
Defining
\begin{equation}
 H(\chi)=3\chi^4+4\chi^3+\chi^2-\chi_c^4,
\end{equation}
one has $H(0)<0$, $H(\chi_c)>0$, and $H'(\chi)>0$ for $\chi>0$.
Hence there is a unique positive extremal root satisfying
\begin{equation}
 \chi_e<\chi_c .
\end{equation}
The critical saddle is therefore separated from the physical endpoint by a finite distance independent of $g$.

All assumptions of Proposition~\ref{prop:saddleorder} are thus satisfied locally.
Since the critical mode has $p=4$, its normalized contribution is
\begin{equation}
 \frac{\Fav-F_{K,c}}{T_c}
 =
 \frac14+\mathcal O(g^{1/2}),
 \label{eq:kerrquarter}
\end{equation}
where $F_{K,c}$ is the classical critical free energy and $T_c=\tau_c/L$.
The quarter-temperature term is therefore not specific to the static charged equation of state, but follows from the quartic order of the rotating critical mode.

To describe the neighborhood of the critical point, set
\begin{equation}
 \Delta\tau=\tau-\tau_c,
 \qquad
 \Delta j=j-j_c .
\end{equation}
We write $\Delta$ for their common order.
Before removing the control-dependent cubic term, the local action contains linear, quadratic, and cubic perturbations around $\sigma_c$.
An analytic translation
\begin{equation}
 \sigma-\sigma_c=y+\mathcal O(\Delta j)
\end{equation}
removes the leading cubic term without changing the linear control map at first order.
The resulting action takes the cusp normal form
\begin{equation}
 \begin{aligned}
 \mathcal I_K
 &=
 \mathcal I_{K,0}(\tau,j)
 -w_Ky+u_Ky^2+\widehat a_{4,K}y^4\\
 &\quad
 +\mathcal O\left(y^5,\Delta y^3,\Delta^2y\right),
 \end{aligned}
 \label{eq:kerrnormalform}
\end{equation}
where
\begin{equation}
 w_K
 =
 \frac{\Delta\tau-m_{\sigma j,c}\Delta j}{\tau_c},
 \label{eq:kerruwdef}
\end{equation}
and
\begin{equation}
 u_K
 =
 \frac{m_{\sigma\sigma j,c}}{2\tau_c}\Delta j .
 \label{eq:kerruanalytic}
\end{equation}
At the critical point,
\begin{equation}
 m_{\sigma j,c}<0,
 \qquad
 m_{\sigma\sigma j,c}>0,
\end{equation}
so the map from $(\Delta\tau,\Delta j)$ to $(w_K,u_K)$ is nonsingular.
Temperature and angular momentum therefore provide independent physical controls of the tilt and curvature of the local quartic potential.

The zero-field tangent is defined by $w_K=0$, which gives
\begin{equation}
 \Delta\tau=m_{\sigma j,c}\Delta j .
 \label{eq:kerrzerofieldpath}
\end{equation}
Since $m_{\sigma\sigma j,c}>0$, the side $\Delta j>0$ has $u_K>0$ and therefore belongs to the one-well sector.
Introducing
\begin{equation}
 \lambda_K
 =
 \frac{u_K}{\sqrt{\widehat a_{4,K}g}},
 \qquad
 h_K
 =
 \frac{w_K}{
 \widehat a_{4,K}^{1/4}g^{3/4}
 },
 \label{eq:kerrscaledfields}
\end{equation}
reduces Eq.~\eqref{eq:kerrnormalform} to the same universal potential
\begin{equation}
 V_K=x^4+\lambda_Kx^2-h_Kx
\end{equation}
that appeared in the RN-AdS analysis.

Along the zero-field one-well direction, $h_K=0$ and $\lambda_K\propto\Delta j/\sqrt g$.
The Kerr-AdS critical window therefore has
\begin{equation}
 \Delta j=\mathcal O(g^{1/2})
\end{equation}
along this trajectory.
No new scaling integral is required.
The normalized crossover is governed by the same function $\Phi$ in Eq.~\eqref{eq:phi},
\begin{equation}
 \frac{\Fav-\Fcl}{\TE}
 =
 \Phi(\lambda_K)
 +\text{subleading corrections},
\end{equation}
with
\begin{equation}
 \Phi(0)=\frac14,
 \qquad
 \Phi(\infty)=\frac12 .
\end{equation}
Thus RN-AdS and Kerr-AdS differ in the physical map from their thermodynamic controls to the cusp fields, while the local saddle-order structure and the quartic-to-Gaussian crossover are the same.

\section{Conclusion and discussion}
\label{sec:discussion}

We have established a saddle-order proposition that determines the leading fluctuation contribution to the ensemble-averaged generalized free energy. For a unique stable interior saddle with a smooth nonzero measure density, a leading potential of even order $p$ gives
\begin{equation}
 \Fav-\Fcl=\frac{\TE}{p}+\mathcal O(\TE g^{2/p}).
\end{equation}
The same order fixes the collective fluctuation width through $g^{1/p}$. The Gaussian half-temperature correction and the quartic quarter-temperature contribution therefore belong to a common semiclassical expansion. At an ordinary critical point, the quadratic restoring term vanishes, and the quartic term determines both the broader fluctuation distribution and its normalized contribution.

The RN-AdS and Kerr-AdS realizations establish the black hole input required by the proposition. In each ensemble, we identify the overall semiclassical scaling of the action, derive a positive quartic coefficient, and verify that the energy measure is regular at a critical saddle separated from the physical endpoint. These analytic checks yield the quarter-temperature contribution with a controlled error of order $g^{1/2}$. The equation of state determines the critical location, the width coefficient, and the physical control map, while the saddle order fixes the leading normalized contribution.

The control maps extend this result from the critical point to its neighborhood. The local cusp structure also appears in off-shell analyses of black hole criticality~\cite{HaoWang2026}. Here the normalized ensemble average assigns a definite fluctuation contribution to this structure. Temperature and charge in RN-AdS, or temperature and angular momentum in Kerr-AdS, unfold the critical saddle into the same quartic normal form. Along the zero-field one-well direction, its scaling function connects the critical and Gaussian contributions. This function is exact for the normal-form integral and gives the leading double-scaling result for the reduced ensembles. At criticality, the quartic contribution survives as $g\to0$, whereas a fixed positive curvature restores the Gaussian value. The uniform integral describes the intervening regime where that curvature becomes comparable to the critical fluctuation scale.

For several collective directions, weighted homogeneity generalizes the saddle-order rule. The leading normalized contribution is $\TE\sum_i w_i$, where the weights describe the simultaneous scaling of the coupled potential. Factorization into independent modes is therefore unnecessary. Noncritical Gaussian directions retain their usual contributions, while the critical sector is governed by its own leading scaling structure. Appendix~\ref{app:general-polynomial} illustrates how competing polynomial couplings can reorganize the dominant weights and introduce logarithmic factors.

The normalized average retains less measure dependence than the partition function. A smooth density approaches a nonzero constant over the shrinking saddle region, and this constant cancels between the numerator and denominator of the average. Its variation affects the subleading terms. An overall normalization also cancels from $\Fav$, but remains in $-\TE\ln Z$. The universal normalized coefficient consequently does not determine the complete logarithmic correction to the gravitational partition function.

Including the gravitational modes omitted from the reduced ensemble requires a further stability analysis. The connection between thermodynamic stability and Euclidean negative modes has been studied in Refs.~\cite{GrossPerryYaffe1982,Prestidge2000,MonteiroSantos2009}. For a class of black branes, Reall also related local thermodynamic instability to the classical Gregory-Laflamme instability~\cite{Reall2001}. These results give physical motivation for examining the full fluctuation spectrum when extending the present calculation. The stable minimum established here controls the retained collective direction; additional modes must be treated with their appropriate measure and integration contour.

The saddle-order proposition determines the leading fluctuation contribution precisely where the Gaussian approximation fails. Together with the uniform crossover, it explains how the critical correction joins the familiar Gaussian result and why this behavior is shared by charged and rotating black hole ensembles.

\begin{acknowledgements}
Ankit Anand is financially supported by the Institute’s
postdoctoral fellowship at IIT Kanpur. This work is supported by the National Natural Science Foundation of China (Grant No. 12405073) and the Natural Science Foundation of Tianjin (Grant No. 25JCQNJC01920).
\end{acknowledgements}


\appendix 

\section{General polynomial couplings and dominant weighted scaling}\label{app:general-polynomial}

In Sec.~\ref{sec:multidimensional}, we assume that the leading local potential is weighted homogeneous. In this appendix, we discuss what happens when we consider a very general leading local potential. We start with a very general form of potential as
\begin{equation}
V_0\!\left(\lambda^{w_1}\delta^1,\ldots,\lambda^{w_n}\delta^n \right) = \lambda V_0(\bm{\delta}) \ ,
\end{equation}
so that the local saddle contribution is governed by the total weight $W=\sum_i w_i$.  A generic local Taylor expansion, however, may contain monomials of different weighted orders.  We examine the two-variable polynomial
\begin{equation}
V(x,y) = c x^m+a x^n y^o+b y^p \ ,
\label{eq:app-general-V}
\end{equation}
and determine which subset of monomials controls the semiclassical scaling.

Throughout this appendix, we consider a unique stable interior saddle at the origin and assume that the local density is smooth and nonzero
\begin{equation}
\rho(x,y;g) = \rho_0+\mathcal O(x,y) ,
\qquad \rho_0>0 \ .
\label{eq:app-density}
\end{equation}
For the explicit real contour examples, it is sufficient to take $m$ and $p$ even and $b>0$, $c>0$ together with conditions on $a,n,o$ ensuring that the full polynomial is positive away from the origin. In the examples with relevant couplings below, we take $a>0$ and even $n,o$ for simplicity.

Near the saddle, write the reduced Euclidean action as
\begin{equation}
I_E(x,y)
=
I_0+\frac{1}{g}V(x,y)+\cdots ,
\label{eq:app-local-action}
\end{equation}
with $V$ given by Eq.~\eqref{eq:app-general-V}.  The corresponding
source-dependent local integral is
\begin{align}
Z_{\rm loc}(s) &= e^{-sI_0}\mathcal N(g) \int dx\,dy\, \rho(x,y;g) \nonumber\\
&\qquad\times \exp\left[ -\frac{s}{g}\left(cx^m+a x^n y^o+b y^p\right)\right] \ .
\label{eq:app-Z-local}
\end{align}
If the mixed term were absent, the two pure stabilizing monomials would
give the natural semiclassical widths
\begin{equation}
x\sim g^{1/m},
\qquad
y\sim g^{1/p}.
\label{eq:app-naive-widths}
\end{equation}
Thus the preliminary scaling weights are
\begin{equation}
w_x^{(0)}=\frac{1}{m},
\qquad
w_y^{(0)}=\frac{1}{p}.
\label{eq:app-preliminary-weights}
\end{equation}

The weighted degree of the mixed monomial with respect to these weights
is
\begin{equation}
\Delta_{\rm w}
\equiv
\frac{n}{m}+\frac{o}{p}.
\label{eq:app-Delta}
\end{equation}

This single combination determines the role of the coupling.

To see this explicitly, introduce
\begin{equation}
\epsilon\equiv\frac{g}{s}
\end{equation}
and rescale
\begin{equation}
x=
\left(\frac{\epsilon}{c}\right)^{1/m}X,
\qquad
y=
\left(\frac{\epsilon}{b}\right)^{1/p}Y.
\label{eq:app-rescaling}
\end{equation}
The Jacobian is
\begin{equation}
dx\,dy
=
c^{-1/m}b^{-1/p}
\epsilon^{1/m+1/p}
dX\,dY.
\label{eq:app-jacobian}
\end{equation}
The exponent becomes
\begin{align}
\frac{V}{\epsilon}
&=
X^m+Y^p
+
\frac{A}{\epsilon^{1-\Delta_{\rm w}}}
X^nY^o, \quad \text{with}\quad A
\equiv
\frac{a}{c^{n/m}b^{o/p}} \ .
\label{eq:app-scaled-potential}
\end{align}
Eq.~\eqref{eq:app-scaled-potential} gives three distinct regimes:
\begin{equation}
\setlength{\arraycolsep}{3pt}
\begin{array}{lll}
\Delta_{\rm w}>1:
&
\epsilon^{\Delta_{\rm w}-1}\rightarrow0,
&
\text{higher-weight coupling},
\\[4pt]
\Delta_{\rm w}=1:
&
\epsilon^{\Delta_{\rm w}-1}=1,
&
\text{weighted-homogeneous coupling},
\\[4pt]
\Delta_{\rm w}<1:
&
\epsilon^{\Delta_{\rm w}-1}\rightarrow\infty,
&
\text{relevant coupling}.
\end{array}
\label{eq:app-three-regimes}
\end{equation}

\subsection{Higher-weight coupling: $\Delta_{\rm w}>1$}

Suppose first that $\Delta_{\rm w}>1$ and define $\kappa
\equiv \Delta_{\rm w}-1>0.$ The mixed term in Eq.~\eqref{eq:app-scaled-potential} is then suppressed by $\epsilon^\kappa$.  Expanding the exponential gives
\begin{align}
e^{-V/\epsilon}
&=
e^{-X^m-Y^p}
\left[
1-A\epsilon^\kappa X^nY^o+\mathcal O(\epsilon^{2\kappa})
\right].
\label{eq:app-expansion-irrelevant}
\end{align}

At leading order, the integral factorizes as
\begin{align}
C_{m,p}
&\equiv
\int_{-\infty}^{\infty}dX
\int_{-\infty}^{\infty}dY\,
e^{-X^m-Y^p}
\nonumber\\
&=
\left[\frac{2}{m}\Gamma\left(\frac{1}{m}\right)\right]
\left[\frac{2}{p}\Gamma\left(\frac{1}{p}\right)\right]
\nonumber\\
&=
\frac{4}{mp}
\Gamma\left(\frac{1}{m}\right)
\Gamma\left(\frac{1}{p}\right).
\label{eq:app-Cmp}
\end{align}

Keeping the leading value $\rho_0$ of the local density, and use $Z_{\rm loc}(s)$ for this truncated integral. The correction from the mixed monomial then gives
\begin{align}
Z_{\rm loc}(s)
&=e^{-sI_0}\mathcal N(g)\rho_0\,c^{-1/m}b^{-1/p}C_{m,p}\nonumber\\
&\qquad\times\left(\frac{g}{s}\right)^{W_0}
\left[1+O\left((g/s)^\kappa\right)
\right],
\label{eq:app-Z-irrelevant}
\end{align}
where
\begin{equation}
W_0=\frac{1}{m}+\frac{1}{p}.
\label{eq:app-W0}
\end{equation}
For the full integral with a smooth density, local variations of the density also contribute at subleading orders. Thus the remainder in Eq.~\eqref{eq:app-Z-irrelevant} applies to the integral with density $\rho_0$, while the leading exponent $W_0$ is unchanged.

The leading source dependence is therefore
\begin{equation}
Z_{\rm loc}(s)\propto e^{-sI_0}g^{W_0}s^{-W_0}.
\end{equation}
Taking the logarithm,
\begin{align}
\ln Z_{\rm loc}(s)
&=-sI_0-W_0\ln s+W_0\ln g \nonumber \\&
+\text{$s$-independent terms}+o(1).
\end{align}
Using
\begin{equation}
\overline F
=
-T_E
\frac{d}{ds}\ln Z(s)
\bigg|_{s=1},
\end{equation}
we obtain
\begin{equation}
\overline F
=
T_E(I_0+W_0)+o(T_E).
\end{equation}
Since
\begin{equation}
F_{\rm cl}=T_E I_0,
\end{equation}
the fluctuation contribution is
\begin{equation}
\frac{\overline F-F_{\rm cl}}{T_E}=\frac{1}{m}+\frac{1}{p}+o(1).
\label{eq:app-F-irrelevant}
\end{equation}
Thus a mixed monomial whose weighted degree is greater than unity does not modify the leading universal coefficient.

As an example, consider $V=x^6+x^4y^2+y^4$, it is easy to check $\Delta_{\rm w}=\tfrac{7}{6}>1.$ The leading weights are
\begin{equation}
w_x=\frac16, \quad w_y=\frac14 \implies 
W=\frac16+\frac14=\frac{5}{12} \ .
\end{equation}
Furthermore, $\kappa = \Delta_{\rm w}-1 = 1/6$. Hence
\begin{equation}
Z_{\rm loc} \propto g^{5/12}, \qquad \frac{\overline F-F_{\rm cl}}{T_E} =\frac{5}{12}+ \mathcal O(g^{1/6}) \ .
\label{eq:app-example-x6-result}
\end{equation}


\subsection{Relevant coupling: $\Delta_{\rm w}<1$}

The most interesting situation occurs when $\Delta_{\rm w}
<1$. The rescaling based only on $x^m$ and $y^p$ now gives
\begin{equation}
A\epsilon^{\Delta_{\rm w}-1}X^nY^o \ ,
\notag
\end{equation}
whose coefficient diverges as $\epsilon\rightarrow0$.  Thus, the mixed monomial cannot be treated as a perturbation. Instead, it changes the dominant scaling itself.
Let
\begin{equation}
x\sim g^{w_x},
\qquad
y\sim g^{w_y}.
\label{eq:app-general-weights}
\end{equation}
For each monomial to be no larger than order $g$ in the dominant region, the weights must satisfy
\begin{equation}
mw_x\geq1,
\qquad
nw_x+ow_y\geq1,
\qquad
pw_y\geq1.
\label{eq:app-weight-constraints}
\end{equation}
The shrinking volume element behaves as
\begin{equation}
dx\,dy
\sim
g^{w_x+w_y}.
\end{equation}
The dominant local volume is therefore determined by
\begin{equation}
W_\star = \min_{w_x,w_y>0}
\left\{ w_x+w_y\right\}\,,
\label{eq:app-Wstar-min}
\end{equation}
where the weights $w_x,w_y$ satisfies Eq.~\eqref{eq:app-weight-constraints}.

The three monomials correspond to the exponent points
\begin{equation}
A=(m,0),
\qquad
B=(n,o),
\qquad
C=(0,p).
\label{eq:app-exponent-points}
\end{equation}
The condition
\begin{equation}
\frac nm+\frac op=1
\end{equation}
places $B$ on the line segment joining $A$ and $C$.  When
$\Delta_{\rm w}>1$, $B$ lies above that segment and is subleading.
When $\Delta_{\rm w}<1$, $B$ lies below it and splits the dominant
boundary into the two faces $AB$ and $BC$.

This is the elementary two-dimensional version of the Newton-polygon organization of degenerate asymptotic integrals
\cite{Varchenko1976Newton,Greenblatt2010Newton}.

\subsubsection{Case I: Dominant face for $n<o$}
For $n<o$, the relevant dominant face is the one joining $(m,0)$ and $(n,o)$. The corresponding weights satisfy
\begin{equation}
mw_x=1,
\qquad
nw_x+ow_y=1.
\end{equation}
The first equation gives $w_x=1/m$\added{,} and substituting into the second gives $w_y = \tfrac{m-n}{mo}$. Therefore
\begin{equation}
W_\star =\frac1m+\frac{m-n}{mo}=\frac{m+o-n}{mo}.
\label{eq:app-W-left}
\end{equation}

\subsubsection{Case II: Dominant face for $n>o$}
For $n>o,$ the dominant face is instead the one joining $(n,o)$ and $(0,p)$. The weights satisfy
\begin{equation}
nw_x+ow_y=1,
\qquad
pw_y=1.
\end{equation}
Thus, it is easy to check $w_y=1/p$ and $w_x =(p-o)/(np).$ The total weight is
\begin{equation}
W_\star = \frac{p-o}{np}+\frac1p = \frac{n+p-o}{np} \ .
\label{eq:app-W-right}
\end{equation}

\subsubsection{Case III: Degenerate scaling for $n=o$}

When $n=o$ is qualitatively different.  The diagonal passes through the exponent point $(n,n)$ itself.  Equation~\eqref{eq:app-Wstar-min} then gives
\begin{equation}
W_\star=\frac1n \ .
\label{eq:app-W-vertex}
\end{equation}
However, the minimizing scaling is not unique. This
degeneracy gives an additional logarithmic factor in the local
partition function, whose asymptotic form is stated for a symmetric example below.

Combining the three cases, we have
\begin{equation}
W_\star=
\begin{cases}
\displaystyle
\frac1m+\frac{m-n}{mo},
&
\Delta_{\rm w}<1,\quad n<o,
\\[10pt]
\displaystyle
\frac{p-o}{np}+\frac1p,
&
\Delta_{\rm w}<1,\quad n>o,
\\[10pt]
\displaystyle
\frac1n,
&
\Delta_{\rm w}<1,\quad n=o.
\end{cases}
\label{eq:app-W-relevant-summary}
\end{equation}

Before closing, let's consider some examples of relevant couplings. As a first, we take an asymmetric example as 
\begin{equation}
V=x^8+\alpha x^2y^4+y^{12} \ , \qquad \text{with}\qquad \alpha>0 \ .
\end{equation}
Here $\Delta_{\rm w} < 1$ and $n=2<o=4$, so the dominant weights are therefore $w_x=1/8$ and $w_y=3/16$. Thus $W_\star =5/16$ and
\begin{equation}
Z_{\rm loc} \propto g^{5/16}, \qquad \frac{\overline F-F_{\rm cl}}{T_E} \rightarrow \frac5{16} \ .
\end{equation}

The second example is a symmetric relevant example, consider\begin{equation}
V=x^6+\alpha x^2y^2+y^6, \qquad \text{with} \qquad \alpha>0.
\label{eq:app-symmetric-relevant}
\end{equation}
Here $\Delta_{\rm w}<1$ and $n=o=2$. The total weight is $W_\star=1/2$, and the corresponding asymptotic form is
\begin{equation}
Z_{\rm loc} \propto g^{1/2}\ln\frac1g \ .
\end{equation}
The normalized first moment satisfies
\begin{equation}
\frac{\overline F-F_{\rm cl}}{T_E}
=\frac12-\frac1{\ln(1/g)}+\cdots .
\end{equation}
This gives another explicit example in which the limiting coefficient $T_E/2$ is not associated with a Gaussian saddle.

Finally, summarizing the result for the trinomial local potential as in Eq.~\eqref{eq:app-general-V}, the weighted degree $\Delta_{\rm w}$ defined in Eq.~\eqref{eq:app-Delta} organizes the dominant scaling as follows.
\begin{equation}
\begin{array}{c|c|c}
\text{condition}
&
\text{mixed coupling}
&
W_\star
\\ \hline
\Delta_{\rm w}>1
&
\text{higher weighted order}
&
\displaystyle
\frac1m+\frac1p
\\[9pt]
\Delta_{\rm w}=1
&
\text{weighted homogeneous}
&
\displaystyle
\frac1m+\frac1p
\\[9pt]
\Delta_{\rm w}<1,\ n<o
&
\text{relevant}
&
\displaystyle
\frac1m+\frac{m-n}{mo}
\\[11pt]
\Delta_{\rm w}<1,\ n>o
&
\text{relevant}
&
\displaystyle
\frac{p-o}{np}+\frac1p
\\[11pt]
\Delta_{\rm w}<1,\ n=o
&
\text{relevant vertex}
&
\displaystyle
\frac1n
\;
\text{with }\ln(\frac{1}{g})
\end{array}
\notag
\end{equation}
Proposition~2 describes the case in which all retained monomials have the same weighted degree.
Couplings of higher-weighted order modify only subleading terms, whereas relevant couplings reorganize the principal saddle scaling itself.
The present appendix identifies the dominant weights and the logarithmic case for the two-variable trinomial in Eq.~\eqref{eq:app-general-V}.
More general analytic potentials may be organized by analogous Newton-polyhedron methods, but a theorem for these potentials requires additional nondegeneracy assumptions and is beyond the scope of the black hole realizations considered here.

\bibliographystyle{ref}
\bibliography{ref}

\end{document}